\documentclass[letterpaper,twocolumn,10pt]{article}
\usepackage{usenix2019_v3}

\usepackage{tikz}
\usepackage{amsmath}
\usepackage{duckuments}
\usepackage{enumitem}  

\usepackage{comment}

\usepackage[]{todonotes} 
\newcommand{\vinesh}[2][inline]{\todo[color=blue!20,#1]{\sf \textbf{Vinesh:} #2}}

\usepackage{multicol}

\begin{document}

\date{}

\title{\Large \bf Assessing the Effectiveness of Membership Inference on Generative Music}

\author{
{\rm Kurtis Chow}\\
kurtishc@uci.edu
\and
{\rm Omar Samiullah}\\
osamiull@uci.edu
\and
{\rm Vinesh Sridhar}\\
vineshs1@uci.edu
\and
{\rm Hewen Zhang}\\
hewenz1@uci.edu
} 

\maketitle

\begin{abstract}
Generative AI systems are quickly improving, now able to produce believable output in several modalities including images, text, and audio. However, this fast development has prompted increased scrutiny concerning user privacy and the use of copyrighted works in training. A recent attack on machine-learning models called \emph{membership inference} lies at the crossroads of these two concerns. The attack is given as input a set of records and a trained model and seeks to identify which of those records may have been used to train the model. On one hand, this attack can be used to identify user data used to train a model, which may violate their privacy especially in sensitive applications such as models trained on medical data. On the other hand, this attack can be used by rights-holders as evidence that a company used their works without permission to train a model. 

Remarkably, it appears that no work has studied the effect of membership inference attacks (MIA) on generative music. Given that the music industry is worth billions of dollars and artists would stand to gain from being able to determine if their works were being used without permission, we believe this is a pressing issue to study. As such, in this work we begin a preliminary study into whether MIAs are effective on generative music. We study the effect of several existing attacks on MuseGAN, a popular and influential generative music model. Similar to prior work on generative audio MIAs, our findings suggest that music data is fairly resilient to known membership inference techniques. 
\end{abstract}

\section{Introduction}

\subsection{Membership Inference and Generative Music}


Recent advances in deep learning technologies have allowed them to produce high-quality and realistic samples of images, text, and audio. Indeed, generative AI has the opportunity to democratize access to tools that aid in productivity, education, art creation and much more. On the other hand, this rise has led to greater scrutiny into the ethics of developing and training large deep learning models. Companies that produce deep learning models have a habit of taking human-created data without requesting permission beforehand. This has led to a slew of lawsuits in the past few years wherein plaintiffs such as Reuters, Universal Media Group, Sony, Getty Images and more allege that AI companies have stolen millions of copyright-protected works in order to build products designed to replace those works~\cite{wired}. 

Furthermore, many have pointed to the privacy risks involved when sensitive data is used to train AI models. See for example, a 2019 Guardian article which alleges that contractors for Apple that manually grade the Siri voice assistant's speech recognition model were routinely exposed to sensitive user audio~\cite{guardianapple}. As recently as January 2025, Apple was forced to pay \$95 Million to settle a lawsuit concerning these privacy violations~\cite{reutersapple}. Other privacy risks involve inadvertently using audio data from children, in violation of the Children’s Online Privacy Protection Act~\cite{miao2019audio}. See also~\cite{kaissis2020secure} for a discussion of privacy and machine learning in medicine. 

A recent attack that straddles both of these issues is called \emph{membership inference}. In a membership inference attack (MIA), the attacker exploits a trained model in order to test if a certain piece of data was used during the training of that model. In this way, the attacker can determine whether their or others' data was stolen or misused when training a deep learning model. 

In this paper, we are concerned with studying the effectiveness of MIAs on generative music tasks. As we discuss in more detail below, most membership inference attacks focus on image tasks. Some recent work also covers audio classification tasks, such as speech recognition, but to the best of our knowledge only one work by Kong {\it et al.}~\cite{kong2023efficient} discusses membership inference on generative audio. Nonetheless, we are the first to study the effect of membership inference attacks on generative music and the first to study generative audio from GANs.

Kong {\it et al.} focuses on text-to-speech generation using diffusion models whereas we consider MIAs on audio-to-audio music generation via GANs. In general, much work in generative AI has moved to diffusion models. However, we believe that our study of GANs remains relevant as the adversarial loss function is still a common technique used in ML-enhanced audio codecs such as SoundStream and \textsc{EnCodec}~\cite{zeghidour2021soundstream,defossez2022high}. These codecs are critical components to recent music-generation models such as MusicLM~\cite{agostinelli2023musiclm}, \textsc{MusicGen}~\cite{copet2023simple}, and MeLoDy~\cite{lam2023efficient}. 

Using our novel research direction, we wish to verify whether membership inference attacks are feasible in this domain. Kong {\it et al.}~\cite{kong2023efficient} found that generative text-to-speech models that produced audio were more resilient to MIAs than text-to-image models. However, we observe that music-generation models tend to overfit their data. See for example, MusicLM~\cite{agostinelli2023musiclm}, where the authors found that around 10\% of certain outputs bore large resemblance to training data. We aim to determine whether the results of Kong {\it et al.} generalize or if different audio modalities produce different behavior under membership inference attacks. 

\subsection{Related Work}

\begin{figure*}
    \centering
    \includegraphics[width=0.6\linewidth]{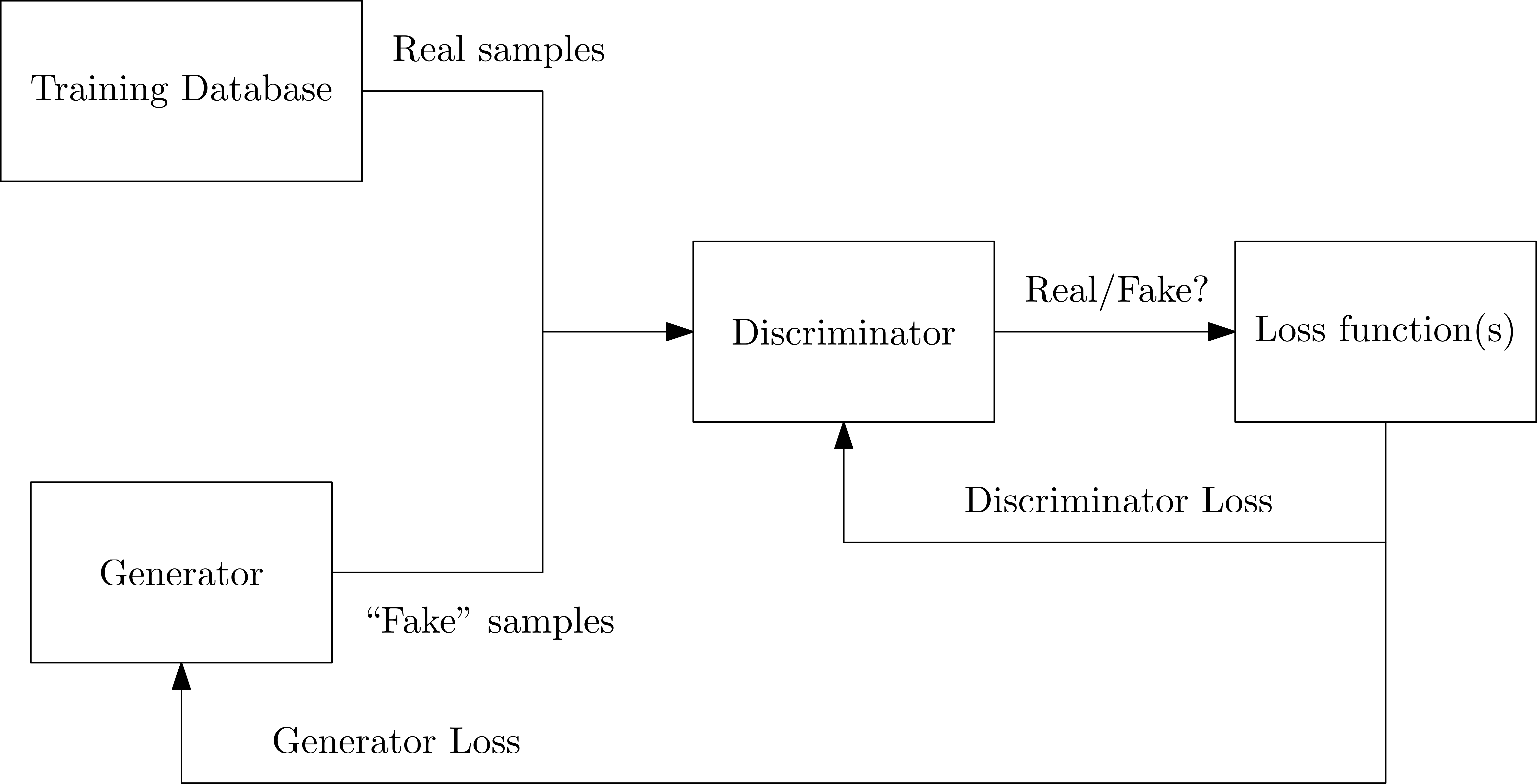}
    \caption{An example of a GAN. A GAN is composed of two main actors, the generator and discriminator. The generator's goal is to produce samples that are similar to the training database and the discriminator's goal is to accurately distinguish between them.}
    \label{fig:GAN-ex}
\end{figure*}

The first membership inference attack was developed in 2017 by Shokri, Stronati, Song, and Shmatikov~\cite{shokri2017membership}. This work influenced all following MIAs, so we review it in detail. In their threat model, they assume access to the structure of the target model (a classifier, in their case) and the distribution of training data, $\mathcal D$. To recover information about the specific records used to train the target model, they invented the \emph{shadow training} technique. The attacker trains several \emph{shadow models} of identical structure to the target model using training and test data sampled from $\mathcal D$. Lastly, the attacker constructs a binary classifier that analyzes the behavior of the shadow models. 

Each shadow model outputs a prediction score when fed a record. The binary classifier is trained on these record, prediction score pairs with the goal of distinguishing between a record whose prediction information indicates it was part of the training set or part of the test set. In a loose sense, the binary classifier is trying to learn how the shadow models overfit their training data. Given enough trained shadow models, the binary classifier can be made accurate enough to predict whether a given record was part of the training set of the target model or not, completing the attack.

The authors consider both the \emph{black-box} setting, where only the target model's final predictions are made available and the \emph{white-box} setting, wherein they have access to intermediate outputs (e.g., gradients) of the target model. The above description covers the black-box setting. In the white-box setting, Shokri {\it et al.}~\cite{shokri2017membership} apply the same attack, but utilize intermediate outputs to increase the richness of training data for the binary classifier, improving accuracy. 

In addition to binary-classifier MIAs, later work has also explored quantitative, heuristics-based approaches. Yeom {\it et al.}~\cite{yeom2018privacy} observe that model correctness and prediction loss can be used to infer membership information. Salem {\it et al.}~\cite{salem2018ml} analyze prediction confidence and prediction entropy in the same vein. See also, e.g.,~\cite{hayes2017logan,song2019auditing,carlini2021extracting,shi2023detecting,choquette2021label,ye2022enhanced,watson2021importance,liu2022membership}.

Hayes, Melis, Danezis, and De Cristofaro~\cite{hayes2017logan} consider the first MIA on generative models. In particular, they attack Generative Adversarial Networks (GANs). A GAN is composed of two components, the generator and the discriminator. The generator turns random noise into a sample that is ideally similar to the training samples taken from $\mathcal D$. The discriminator's job is to determine whether a sample came from the training set or from the generator. 
See Figure~\ref{fig:GAN-ex}.
As such, in a white-box setting, we can simply feed our samples into the discriminator. The discriminator outputs a value that indicates its prediction confidence. Using the heuristics-based methods from above, we can use this value to predict whether a given sample is from the original training set. 

In the black-box setting, however, we do not have direct access to the discriminator. Interestingly, the authors find that in GANs, Shokri {\it et al.}'s shadow training approach does not work~\cite{shokri2017membership}. As such, Hayes {\it et al.} propose a weaker black-box attack that assumes the adversary knows certain samples used to train the target model~\cite{hayes2017logan}. 

Other works that study MIAs on GANs include~\cite{hilprecht2019monte,liu2019performing,chen2020gan,chen2021gan,lin2021privacy,zhang2024generated}. Recent work has also started to explore MIAs on diffusion models~\cite{zhang2024generated,duan2023diffusion,li2024towards,kong2023efficient,matsumoto2023membership,pang2023white}.

We observe that the works stated above primarily focus on image and text modalities. To the best of our knowledge, the first work that studies MIAs on models trained on audio data is a 2019 paper by Miao {\it et al.}~\cite{miao2019audio}. The authors of this work target speech recognition models, which are classification models. 
Accordingly, they employ the original shadow training methods of Shokri {\it et al.}~\cite{shokri2017membership} to determine if the same techniques generalize to audio. Miao {\it et al.}~\cite{miao2019audio} find that they in fact do, with inference accuracy significantly above random chance in most of their tests. Several works have since verified~\cite{Shah2021,feng2023review} and extended~\cite{chen2023slmia,teixeira2024improving,cheng2024gibberishneedmembershipinference,leschanowsky2024examining,tseng2022membershipinferenceattacksselfsupervised,miao2022no} their results to similar audio classification models. 

In contrast, we consider membership inference on generative audio, specifically generative music. As described in the previous section, the work most closely aligned with ours is by Kong {\it et al.}~\cite{kong2023efficient}. They consider MIAs on diffusion models that provide generative text-to-speech. In contrast, we study MIAs on generative music constructed with GANs. Despite both being audio, we believe that speech data is much more uniform than music data, which may change the effectiveness of existing MIAs on this specific modality. Furthermore, their threat model assumes ``gray-box'' access to intermediate output of the model, whereas we consider both white-box and fully black-box approaches.

\section{Preliminaries}

\subsection{MuseGAN}

MuseGAN~\cite{dong2017museganmultitracksequentialgenerative} is a GAN-based framework for symbolic multi-track music generation. Unlike audio waveform models, MuseGAN operates on discrete symbolic data (e.g., MIDI files encoded into numpy arrays), and generates multiple instrument tracks (e.g., drums, bass, piano as numpy arrays) simultaneously to ensure harmonic and rhythmic coherence. 

It adopts a standard GAN Figure~\ref{fig:GAN-ex} setup with a \textit{generator} that maps a latent vector to multi-track sequences, and a \textit{discriminator} that distinguishes real from generated samples. To handle both temporal and inter-track dependencies, MuseGAN proposes several variants:

\begin{itemize}[leftmargin=*]
    \item \textbf{Baseline GAN:} Generates tracks independently without modeling inter-track relations.
    \item \textbf{Jamming GAN:} Uses a shared latent vector but generates tracks separately to promote partial coordination.
    \item \textbf{Composer GAN:} Adds a shared latent space with individual generators for nuanced track-level generation.
    \item \textbf{Full MuseGAN:} Models both track-wise and sequential structures jointly, yielding coherent, bar-level multi-track outputs.
\end{itemize}

We have chosen MuseGAN because it is one of the older GAN-based music-generation models and one of the most popular. This is made clear by the fact that it is cited by and has influenced many modern music-generation models such as MusicGen~\cite{copet2023simple}, EMOPIA~\cite{hung2021emopia}, Popmag~\cite{ren2020popmag}, and Ditto~\cite{novack2024ditto,novack2024ditto2}. 

We analyze the Composer variant of MuseGAN, a deep generative framework designed for symbolic multi-track music generation. 
We do so because it appears to be the most similar to standard GANs and so would be most likely to be susceptible to the same kinds of membership inference attacks.


\begin{figure*}
    \centering
    \includegraphics[width=0.7\linewidth]{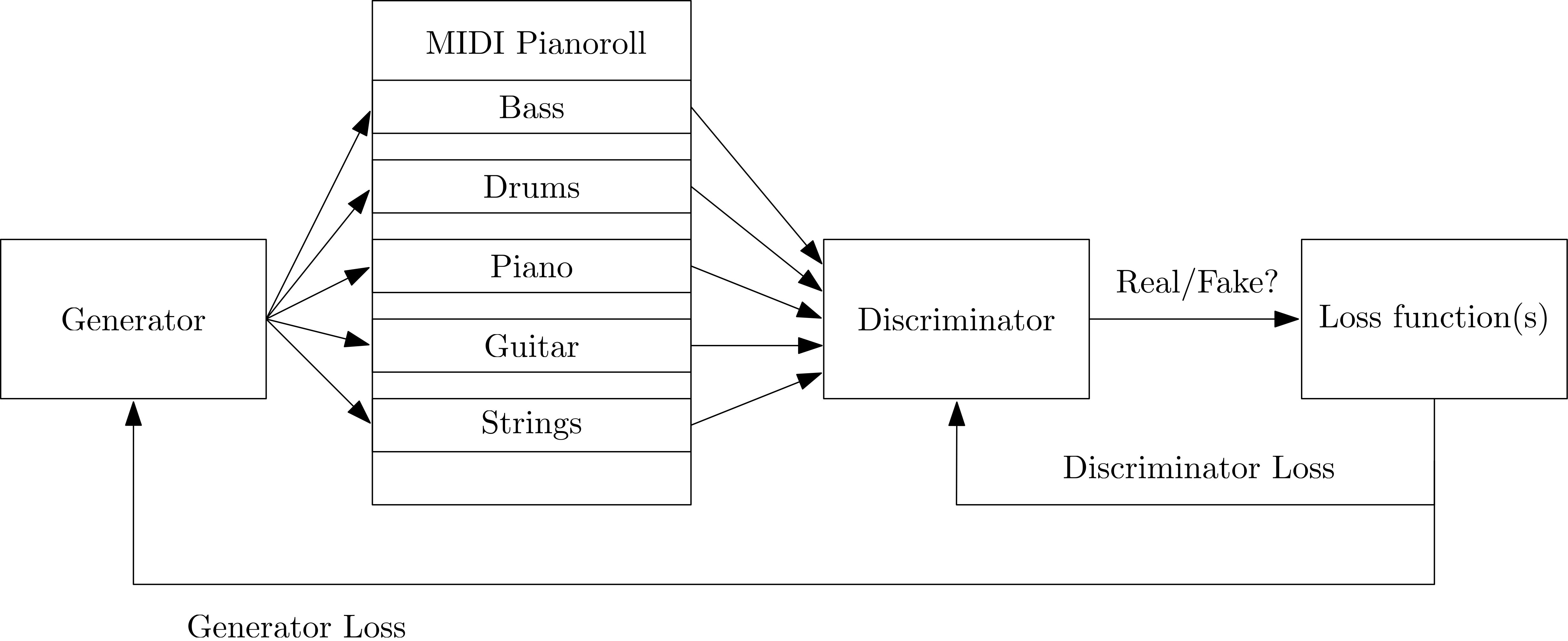}
    \caption{A simplified depiction of the MuseGAN composer model, adapted from~\cite{dong2018musegan}. Their main contribution is to preprocess music 5-track pianorolls such that they can be input into the generator and discriminator while also being able to be recovered as MIDI files. Care is made to ensure the five tracks cohere musically. To this end, they introduce harmony-based methods to measure loss.}
    \label{fig:MuseGAN-ex}
\end{figure*}

\subsection{White-box Discriminator Attack}

The LOGAN framework~\cite{hayes2017logan} introduces a white-box membership inference attack that leverages the discriminator component of a GAN. Since discriminators are trained to distinguish between real and generated data, they often overfit to training samples and yield higher confidence scores for members of the training set. In the white-box setting, an adversary has access to the internal parameters and architecture of the discriminator $D$. The attacker inputs a set of candidate data samples into the discriminator and collects the output confidence scores. By ranking the samples according to these scores, the top-$n$ highest-scoring instances are inferred to be members of the training set ($n$ is the number of training samples). This attack is simple yet powerful: it requires no additional training and only assumes access to the discriminator’s internals. Even when the generator generalizes well, the discriminator may still leak membership information, exposing privacy vulnerabilities that are specific to the adversarial training process.

\subsection{Black-box Monte Carlo Attack}
The Monte Carlo (MC) attack~\cite{hilprecht2019monte} is a black-box membership inference method that relies on output similarity to infer whether a target sample was part of a generative model's training set. The attacker repeatedly queries the generator to produce synthetic samples and compares each to the target instance using similarity metrics such as Euclidean distance or more advanced measures like PCA-reduced distances, HOG features, or color histograms. If many generated samples are close to the target under the chosen metric, the sample is inferred to be a likely training member. Notably, this approach requires no access to internal model parameters or gradients, making it effective in deployed, API-based settings.

\subsubsection{SAP Variation}
The Monte Carlo method proposed by Hilprecht \textit{et al.} ~\cite{hilprecht2019monte} generates a large number (upwards of a million) of samples. These samples are then each compared against hundreds of training and tests samples. The computation load of this attack on our music data was too high for our resources. 
As a result, we turned to the Monte Carlo membership inference attack implemented by SAP, a well-known consultancy company~\cite{sap_attack_py}. 
Their solution instead
generates many random samples from a stash of pre-generated samples in order to balance computational cost and effectiveness.

\section{Methodology}
\subsection{Approach}

To evaluate the success of MIAs on music GANs, we used MuseGAN as a target model. We also implemented two attacks: white-box discriminator and Monte Carlo MIAs. These attacks aims to test both the traditional strategy of MIAs(leveraging the discriminator) and a class of black-box attacks, namely Monte Carlo MIAs.

Based on Hayes {\it et al.}\cite{hayes2017logan}, we expect the white-box discriminator MIA to achieve a high success rate(95\% in their work). However, it should be slightly lower as Kong {\it et al.}~\cite{kong2023efficient} suggests audio data is more resilient to MIAs. Monte Carlo MIAs should perform similarly but have less success than the white-box attack because they are a black-box attack. A large component of these well-performing expectations is that since generative music models tend to produce samples similar to their training data, these attacks are expected to perform well in spite of music data's general resilience.

To further explore generative music data's resilience against MIAs, we trained two MuseGAN models. A model, labeled $M_{default}$, was trained and tested on a 50:50 split of the original data set. This model was used as a benchmark for evaluation. For comparison, we trained another model on only a tenth of the data set. This model, $M_{overfitted}$, weighs the results in favor of MIAs because they perform better when targeting models that are over-fitted. In turn, both attacks should perform better than their counterparts when targeting this model.

\subsection{Implementations}

Our attacks were implemented using a python notebook provided by the original authors of MuseGAN. This version was implemented in PyTorch unlike the paper's TensorFlow GitHub. The attacks were implemented in PyTorch for compatibility with the model.

We originally used the pre-trained model provided with the Tensorflow implementation, but this was built on Tensorflow version 1, leading to several compatibility issues. The code itself had its own errors due to several changes upon a variety of requests\cite{salu133445_2023}. These issues led us to instead use their PyTorch implementation. No pre-trained model is provided by the PyTorch implementation leading to significant limitations in training due to strict computational constraints.

The data set used for training was the Lakh Pianoroll Dataset(LPD)\cite{dong2018musegan}. More specifically, a subset called LPD-5 Cleansed where it is restricted to well-formatted data shaped into five tracks(Drums, Strings, Piano, Guitar, Bass). This decision is to align with the paper. LPD-5 Cleansed consists of 21,425 five-track pianorolls.

Both models $M_{default}$ and $M_{overfitted}$ differ in their training data set and in training rounds. $M_{default}$ was trained on fifty percent of LPD-5 Cleansed randomly selected. Selection was done on the tracks' labels, leading to the number of generated samples for training being different. For $M_{default}$ 12,992 samples were made from the training set while its test set consists of 13,162 samples. Meanwhile, $M_{overfitted}$ was trained on ten percent of LPD-5 Cleansed. This leads to 2,639 samples for training and 23,515 for testing. $M_{default}$ was trained for 20,000 rounds while $M_{overfitted}$ was trained for 200,000 rounds, ten times more to overfit $M_{overfitted}$. Lastly, $M_{default}$ was saved every 1,000 rounds while $M_{overfitted}$ was saved every 20,000 rounds.

Training was performed on a variety of platforms due to time and computational limitations. Training of $M_{default}$ was done on an under-clocked NVIDIA 4070 GPU with 8GB of VRAM. Training of $M_{overfitted}$ was done on a server equipped with a NVIDIA 4090 GPU due to the memory challenges with the extra training rounds. Testing was done on Google Colab's free CPU run-times. This was made possible by saving the models periodically during training and inference generally being less computational intensive than training.

\subsection{White-Box/Discriminator Attack}

The white-box attack consisted of loading our trained models' discriminator and having them predict if a sample was from the training or test set. MuseGAN by default does not provide predictions as a label or percentage. Instead, it is a score where larger values indicate the discriminator is confident the sample came from the training set.

For both models, multiple checkpoints were taken in order to perform the attack over their training iterations. The attack first loads the saved model's discriminator and is given each sample one-by-one to predict a score. We then sort them, labeling the top-scoring $N$ samples as predicted to be from the training set where $N$ is the number of training samples used. From these predictions, we collect the true and false positives and negatives for analysis.

\subsection{Black-box Monte Carlo Attack}
The core idea of MC attack is that, if a generative model has overfitted to its training set, then outputs will fall "closer" to training samples significantly than to unseen test samples. According to \cite{hilprecht2019monte}, the basic MC membership score is given by
\begin{equation}
\widehat{f}_{\mathrm{MC}\!-\varepsilon}(x)
\;=\;
\frac{1}{n} \sum_{i=1}^{n} 
\mathbf{1}\bigl[\,d(g_{i},\,x)\le \varepsilon \bigr],
\end{equation}
where $g_i$ are generated samples, $d(g_i,\,x)$ is a distance metric, and $\varepsilon$ is a threshold that decided by some heuristic. The authors use heuristics based on the median distance in the data set or some top percentile of distances.
In the image-based MC attack, distance measures including Principal Component Analysis(PCA), Histogram of Oriented Gradients(HOG), and Color Histogram(CHIST) output feature vectors which are used to calculate Euclidean distances for distance metric. Additionally, MC attack is implemented in 2 ways, Single MI and Set MI. $M$ records are drawn from training data, and another $M$ records are drawn from test data respectively. The distances between $2M$ records and generated samples from the generative model are calculated. For the Single MI attack, the top distances $M$ are selected and the accuracy of this attack type is the proportion of training data in this set. For the Set MI attack, top $M$ of the distances are labeled as set $\mathcal{R}$. The set that has more records in $\mathcal{R}$ will be labeled as training set.

\subsection{Metrics}
For the white box discriminator attack. we measured its success rate, accuracy, precision, recall/true positive rate, false positive rate, and F1-scores. Success rate was calculated as the number of correct predictions of being in the training set over the real number of training set samples. These measurements were taken at each saved training step per model. For $M_{default}$, every 1,000 steps and for $M_{overfitted}$, every 20,000 steps.

In the black box MC attack, considering the different features of audio signal from image data, Euclidean distances on raw data were used to perform the attack. To calculate and evaluate MC metrics, the GitHub code associated with~\cite{hilprecht2019monte} was adopted as reference. Generated samples by different model rounds ranging from 20000 to 200000 were used to compare attack performance. We tested the attack accuracies based on training data and test data in $M_{overfitted}$. The test data was randomly selected from the original test data to ensure that it was the same size as the training data. Several heuristics were also applied, including $1\%$, $0.1\%$, $0.01\%$ and the median heuristic.

\section{Evaluation}
\begin{table*}
\centering
\begin{tabular}{lrrrrrr}
\hline
Iterations & Success Rate & Accuracy & Precision & Recall/TPR & FPR & F1 \\
\hline
1000 & 0.505 & 0.508 & 0.505 & 0.505 & 0.488 & 0.505 \\
2000 & 0.504 & 0.508 & 0.505 & 0.504 & 0.489 & 0.504 \\
3000 & 0.506 & 0.509 & 0.506 & 0.506 & 0.488 & 0.506 \\
4000 & 0.503 & 0.506 & 0.503 & 0.503 & 0.491 & 0.503 \\
5000 & 0.503 & 0.506 & 0.503 & 0.503 & 0.491 & 0.503 \\
6000 & 0.499 & 0.502 & 0.499 & 0.499 & 0.494 & 0.499 \\
7000 & 0.507 & 0.511 & 0.507 & 0.507 & 0.486 & 0.507 \\
8000 & 0.503 & 0.507 & 0.503 & 0.503 & 0.490 & 0.503 \\
9000 & 0.500 & 0.504 & 0.501 & 0.500 & 0.493 & 0.500 \\
10000 & 0.502 & 0.505 & 0.502 & 0.502 & 0.492 & 0.502 \\
11000 & 0.502 & 0.505 & 0.502 & 0.502 & 0.492 & 0.502 \\
12000 & 0.506 & 0.509 & 0.506 & 0.506 & 0.488 & 0.506 \\
13000 & 0.505 & 0.509 & 0.505 & 0.505 & 0.488 & 0.505 \\
14000 & 0.504 & 0.507 & 0.504 & 0.504 & 0.490 & 0.504 \\
15000 & 0.511 & 0.514 & 0.511 & 0.511 & 0.483 & 0.511 \\
16000 & 0.503 & 0.506 & 0.503 & 0.503 & 0.491 & 0.503 \\
17000 & 0.512 & 0.515 & 0.512 & 0.512 & 0.482 & 0.512 \\
18000 & 0.497 & 0.500 & 0.497 & 0.497 & 0.497 & 0.497 \\
19000 & 0.496 & 0.499 & 0.496 & 0.496 & 0.498 & 0.496 \\
20000 & 0.504 & 0.507 & 0.504 & 0.504 & 0.490 & 0.504 \\
\hline
\end{tabular}
\caption{White-Box Attack Metrics Across Training of $M_{default}$}
\label{wb_default_table}
\end{table*}
\begin{table*}
\centering
\begin{tabular}{lrrrrrr}
\hline
Iterations & Success Rate & Accuracy & Precision & Recall/TPR & FPR & F1 \\
\hline
20000 & 0.121 & 0.823 & 0.121 & 0.121 & 0.099 & 0.121 \\
40000 & 0.102 & 0.819 & 0.102 & 0.102 & 0.101 & 0.102 \\
60000 & 0.106 & 0.820 & 0.107 & 0.106 & 0.100 & 0.106 \\
80000 & 0.134 & 0.825 & 0.134 & 0.134 & 0.097 & 0.134 \\
100000 & 0.116 & 0.822 & 0.116 & 0.116 & 0.099 & 0.116 \\
120000 & 0.103 & 0.819 & 0.103 & 0.103 & 0.101 & 0.103 \\
140000 & 0.103 & 0.819 & 0.103 & 0.103 & 0.101 & 0.103 \\
160000 & 0.115 & 0.821 & 0.115 & 0.115 & 0.099 & 0.115 \\
180000 & 0.099 & 0.818 & 0.099 & 0.099 & 0.101 & 0.099 \\
200000 & 0.108 & 0.820 & 0.108 & 0.108 & 0.100 & 0.108 \\
\hline
\end{tabular}
\caption{White-Box Attack Metrics Across Training of $M_{overfitted}$}
\label{wb_overfitted_table}
\end{table*}
\begin{table*}
\centering
\begin{tabular}{lrr}
\hline
Epochs & Single MI Accuracy & Set MI Accuracy \\
\hline
20000 & 0.501 & 1.000 \\
40000 & 0.504 & 1.000 \\
60000 & 0.509 & 1.000 \\
80000 & 0.502 & 1.000 \\
100000 & 0.503 & 1.000 \\
120000 & 0.505 & 1.000 \\
140000 & 0.503 & 1.000 \\
160000 & 0.503 & 1.000 \\
180000 & 0.504 & 1.000 \\
200000 & 0.504 & 1.000 \\
\hline
\end{tabular}
\caption{Monte Carlo attack accuracy based on median heuristic.}
\label{MC}
\end{table*}
\begin{figure}
    \centering
    \includegraphics[width=0.8\linewidth]{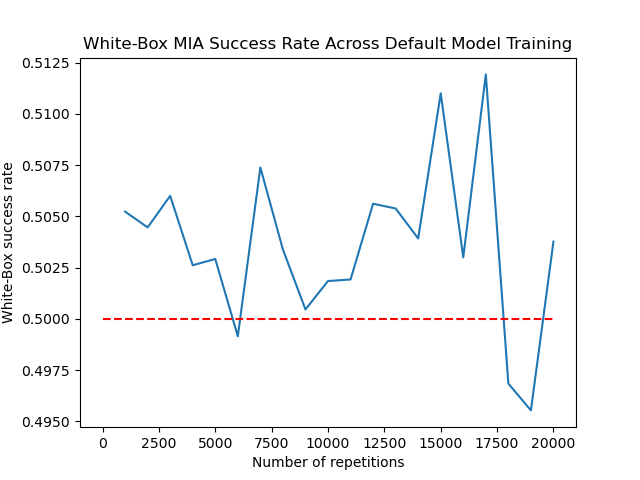}
    \includegraphics[width=0.8\linewidth]{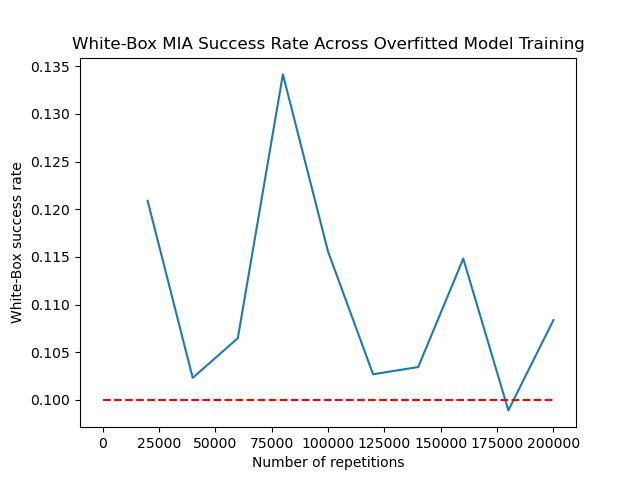}
    \caption{Attack Success Rate Across Training}
    \label{Attack Success Rate Across Training}
\end{figure}

In this section, we will present the findings from our attacks. Overall, both attacks performed poorly and were unable to effectively distinguish if a sample is from the training set.

For $M_{default}$, the white-box attack was most successful when it was trained for 17000 rounds. However, its average success rate was 50.36\%. This indicates the attack was no better than guessing. In addition, the average precision and recall were roughly the same, further cementing the previous point. The average false positive rate was around 49\%, lower than the other metrics, but not significantly(Table \ref{wb_default_table}).

For $M_{overfitted}$, we expected the results to be more successful because an over-fitted model should lead to a discriminator who would be better at distinguishing training samples. However, the attack ultimately achieved even worse performance. It's average success rate was 11\%, peaking at 13.4\% at 80,000 rounds of training(Figure \ref{Attack Success Rate Across Training}). The other metrics reflect this reality as well(Table \ref{wb_overfitted_table}), being around 10\%. Notably, it had a high overall accuracy because it predicted most test samples were from the test set. This suggests the over-fitted model may be overly cautious on if a sample is from the training set.

The results of the white-box discriminator attack suggests that music data is incredibly resilient to typical membership inference attacks. While shadow model training was not done due to constraints, given that they appear to perform worse than their white-box counterpart\cite{hayes2017logan}, we expect similar results. It appears that generative music GANs, or at least MuseGAN, are resilient to MIAs using discriminators.

In \cite{hilprecht2019monte}'s original work on MC membership inference against image-based GANs, Single MI accuracies of GANs often exceed $50-60\%$ on standard datasets like MNIST and CIFAR-10, and rise when models overfit. Set MI obtains better performance and the accuracies vary from $50-80\%$ based on different heuristics and datasets. We therefore anticipated a trend that the accuracies rise with epoch count increasing.

Across all checkpoints from 20000 to 200000 rounds and regardless of whether we chose median heuristic or fixed percentiles (1\%, 0. 1\%, 0. 01\%), the single MI MC attack consistently achieved about 48\%-51\% accuracy, which is essentially random guessing. In addition, the results did not show an upward trend as training progressed. As Table~\ref{MC} indicated, Single MI accuracy had no improvement when models overfitted, meaning the attack cannot distinguish whether an individual sample is from training set. This revealed that applying Euclidean distance on raw pianoroll music data failed to capture meaningful membership information. In contrast, Set MI attack always achieved 100\% accuracy in all configurations. In the current implementation, Set MI accuracy was computed exactly once by comparing the total distances based on training samples against the total distances based on test samples, which was completely derived from the GitHub code of the original MC attack. As a result, the accuracy would be either 0\% or 100\%. It did not represent an average success rate over multiple experiments or random subsets, and could only represent one aggregate comparison, which did not significantly verify the practicality of the method.

The basic reason why MC attack did not perform as expected was that, unlike image pixels, symbolic pianorolls occupy a very high‐dimensional, sparse space where Euclidean distance does not reflect perceptual “closeness” to training data. Also, the results were insensitive to heuristics because MuseGAN’s generator did not overfit to individual training samples in Euclidean terms and there simply were not many generated samples around true training samples. In sum, MC attack based on image data is not suited for MC membership inference on music data and it needs further work to explore appropriate distance measurements and metrics algorithms.

\section{Discussion and Future Work}

Our preliminary results seem to suggest that MuseGAN is resistant to membership inference attacks (MIAs) under our training setup. This contrasts with prior work showing success against overfitted generative models in the image and text domain, e.g.,~\cite{chen2020gan,hilprecht2019monte,hayes2017logan}. Both white-box (LOGAN)~\cite{hayes2017logan} and black-box (Monte Carlo)~\cite{hilprecht2019monte} attacks yielded low attack success rates.

Two possible causes are the following. (1) limited overfitting, due to model architecture or computational constraints in training. (2) symbolic music, due to its high structural redundancy and limited semantic range, may make it harder for adversaries to distinguish training samples from non-members.
This aligns with Kong {\it et al.}~\cite{kong2023efficient}, who observed similar robustness in speech-generation models. 
However, we acknowledge that our 
choice of Euclidean distance to determine the similarity of two music tracks may not be 
appropriate for the audio domain
and may be the cause of our attacks' poor performance.

Future work may consider the following directions.

\begin{itemize}
    \item Apply music-centric similarity metrics for the Monte Carlo attack, such as Harte, Sandler, and Gasser's \emph{tonal distance}~\cite{harte2006detecting}.
    \item Test MIAs on other music generation models (e.g., MusicLM~\cite{agostinelli2023musiclm}, PopMAG~\cite{ren2020popmag}) to compare modality-specific risks.
    \item Improve attack efficiency via techniques like importance sampling or model distillation.
    \item Explore the effect of corresponding MIA defenses such as differential privacy and adversarial regularization.
\end{itemize}

These directions aim to clarify privacy vulnerabilities in symbolic generative models and guide safer training practices.

\bibliographystyle{plain}
\bibliography{refs}

\end{document}